\documentclass[11pt,a4paper,usenatbib]{emulateapj}
\slugcomment{{\sc Accepted to ApJ:} 15 November, 2008}

\usepackage{amsmath}

\newcommand{\Msol}{\rm\,M_{\odot}}

\newcommand{\Rc}{\rm\,R_{C}}
\newcommand{\UBVRI}{\rm\,UBVR_{C}I_{C}}
\newcommand{\Gyr}{\rm\,Gyr}
\newcommand{\Msolyr}{\rm\,M_{\odot}\,yr^{-1}}

\newcommand{\sqdegr}{\raisebox{0.65ex}{\tiny\fbox{$ $}}\,$^{\circ}$}
\newcommand{\kmsMpc}{\rm\,km\,s^{-1}\,Mpc^{-1}}
\newcommand{\kms}{\rm\,km\,s^{-1}}

\begin{document}

\title{Morphological Composition of z$\sim$0.4 groups: The site of S0 formation}

\author{D.~J.~Wilman$^{1,5}$,~A.~Oemler~Jr$^2$,~J.~S.~Mulchaey$^2$,~S.~L.~McGee$^3$,~M.~L.~Balogh$^3$,~R.~G.~Bower$^4$}
\affil{$^1$Max-Planck-Institut f\"ur extraterrestrische Physik, Giessenbachstra\ss e, D-85748 Garching, Germany.\\
$^2$Observatories of the Carnegie Institution, 813 Santa Barbara Street, Pasadena, California, U.S.A.\\
$^3$Department of Physics and Astronomy, University of Waterloo, Waterloo, Ontario, N2L 3G1, Canada.\\
$^4$Physics Department, University of Durham, South Road, Durham DH1 3LE, U.K.\\
$^5$email: dwilman@mpe.mpg.de\\
}

\begin{abstract}

The low redshift Universe ($z\lesssim0.5$) is not a dull place.
Processes leading to the suppression of star formation and morphological transformation are prevalent: 
this is particularly evident in the dramatic upturn in the fraction of S0-type galaxies in clusters. 
However, until now, the process and environment of formation has remained unidentified. 
We present a morphological analysis of galaxies in the optically-selected (spectroscopic friends-of-friends) group and field environments at $z\sim0.4$. 
Groups contain a much higher fraction of S0s at fixed luminosity than the lower density field, with $>$99.999\% confidence. 
Indeed the S0 fraction in groups is at least as high as in $z\sim0.4$ clusters and X-ray selected groups, which have more luminous Intra Group Medium (IGM). 
An excess of S0s at $\geq$0.3h$_{75}^{-1}$Mpc from the group centre with respect to the inner regions, existing with 97\% confidence at fixed luminosity, tells us that formation is not restricted to, and possibly even avoids, the group cores. 
Interactions with a bright X-ray emitting IGM cannot be important for the formation of the majority of S0s in the Universe.
In contrast to S0s, the fraction of elliptical galaxies in groups at fixed luminosity is similar to the field, whilst the brightest ellipticals are strongly enhanced towards the group centres ($>$99.999\% confidence within 0.3h$_{75}^{-1}$Mpc). 
Interestingly, whilst spirals are altogether less common in groups than in the field, there is also an excess of faint, Sc$+$ type spirals within 0.3h$_{75}^{-1}$Mpc of the group centres (99.953\% confidence). 
We conclude that the group and sub-group environments must be dominant for the formation of S0 galaxies, 
and that minor mergers, galaxy harassment and tidal interactions are the most likely responsible mechanisms. 
This has implications not only for the inferred pre-processing of cluster galaxies, but also for the global morphological and star formation budget of galaxies: as hierarchical clustering progresses, more galaxies will be subject to these transformations as they enter the group environment.

\end{abstract}


\section{Introduction}\label{sec:intro}

Morphologically early type galaxies (ellipticals and S0s) inhabit dense regions of the Universe such as rich clusters, whilst late type spirals are more common in the field. 
This result dates back to the early days of extragalactic astronomy \citep[e.g.][]{Abell65,Oemler74,Bahcall77} and is still of central importance to our understanding of how galaxies evolve in relation to the large scale structure in which they live.
The composition of the galaxy population in clusters also depends upon redshift: 
\citet{BO84} first showed that the fraction of blue (implying star forming) galaxies in clusters increases with redshift to $z\sim0.5$, indicating an evolving population even in these most extreme environments. 
Morphologically, \citet[][D97]{Dressler97} show that, for bright galaxies at least, the increasing fraction of early-type galaxies since $z\sim0.5$ corresponds mainly to lenticular S0 galaxies, with a roughly constant elliptical fraction. 
This conclusion was strengthened by \citet[][F00]{Fasano00} adding clusters from their own data and from the sample of \citet[][C98]{Couch98} to those of D97. 
They show that the cluster S0 to elliptical ratio is, on average, a factor $\sim5$ higher at $z\sim0$ than at $z\sim0.5$. 
At higher redshift, there is no evidence for any further evolution of the S0 fraction in clusters to $z\sim 1$ \citep{Smith05,Postman05,Desai07}. 
Individual clusters at $z\gtrsim0.2$ exhibit a large scatter in their morphological composition, correlating with level of substructure and concentration, especially of elliptical galaxies, in these clusters (D97,F00). 
Indeed, accounting for the possible bias of the classifier, \citet{Andreon97} and \citet{Fabricant00} discuss individual clusters at intermediate redshift with S0 to elliptical ratios consistent with those at $z\sim0$. 

It is natural to relate the cluster to cluster scatter in galaxy properties and dependence on substructure at intermediate redshift to the formation history of the cluster, which can be quite different within the $\Lambda$CDM paradigm of bottom-up structure growth. 
The presence of strong relations at $z\sim0$ between morphological composition and cluster-centric radius \citep{Melnick77}, and local density \citep[][D80]{Dressler80} imply that by that time the galaxy population has reacted to its environment over the Hubble time in a predictable fashion. 
The morphology-density relation of D80 extends to low concentration clusters and to galaxy groups \citep{Postman84}, and so the integrated evolutionary history seems to be predictable even on these scales. 
Nonetheless, differences from the average morphology-density relation are still evident at $z\sim0$: Relatively massive, evolved X-ray bright groups have a lower spiral fraction than do clusters at similar physical density \citep{Helsdon03}, whilst compact groups have higher spiral fractions \citep{Mamon86,Hickson88} at high projected density. 

To explain the evolution of morphology-density relations and the fraction of S0 galaxies, it is necessary to invoke morphology-changing processes which react to the local environment, and can be active on group scales.
In fact, if S0 formation takes place in groups and/or even lower density environments, then this can have a profound effect on the overall galaxy population, since $\gtrsim 50\%$ of all galaxies are in groups by $z\sim0$ \citep{Eke04}. 
This kind of pre-processing can also drive the evolution of the cluster population. 
For example the evolving fraction of S0s in clusters might result from the evolving population of newly accreted galaxies from infalling groups and the field. 
Similar ideas have been invoked to explain the evolving dependence of the fraction of star forming galaxies on cluster mass \citep{Ellingson01,Poggianti06} and the spectrophotometric properties of S0 and ``passive spiral'' populations in clusters \citep[][although note that the emission line threshold used to define ``passive spirals'' is not enough to ensure a complete lack of star formation \citep{Wilman08}]{Moran07}. 
At intermediate redshift the members of optically selected groups (Wilman et al., 2005a\nocite{Wilman05}) and even more massive X-ray selected groups \citep[][J07]{Jeltema07} vary strongly on a group to group basis, from elliptical/passive galaxy dominated to spiral / star forming galaxy dominated. 
This does not appear to be as well correlated with velocity dispersion or local density as in low redshift groups \citep{Postman84,Zabludoff98,Poggianti06}. 
Therefore it is likely that this is a critical regime and epoch, in which important evolutionary processes have been felt by galaxies in some but not all groups, as a function of their environmental history. 

We have compiled a sample of $0.3\leq z \leq0.55$ groups, selected in redshift space from the CNOC2 (Canadian Network for Observational Cosmology) redshift survey \citep[][Wilman et al., 2005a\nocite{Wilman05}]{Carlberg01}. 
Our multiwavelength dataset includes deep, high resolution Hubble Space Telescope (HST) Advanced Camera for Surveys (ACS) observations of these groups, for which galaxies have been morphologically classified. 
With this dataset it is possible to address directly the importance of S0 formation in galaxy groups at intermediate redshift, and assess their importance in driving the evolving fraction of S0s in clusters via pre-processing. 
This complements \citet{McGee08} in which the bulge-disk decomposition software {\rm GIM2D},\citep{Simard02} was applied to our sample, allowing separate investigations of bulge and disk components and their dependence on the group environment.
Data and sample selection are presented in Section~\ref{sec:data}, including a discussion of the morphological classification process. 
Section~\ref{sec:composition} presents the morphological composition of group and field samples as a function of luminosity, and differences with environment are assessed for statistical significance at fixed luminosity. 
Segregation within groups is then investigated in Section~\ref{sec:radialtrends}, by looking for trends of composition with distance from the centre of groups. 
The morphological composition of our group and field samples are put into context in Section~\ref{sec:disc} by comparing with that seen in other cluster and group samples as a function of redshift. This leads to the conclusion, summarized in Section~\ref{sec:concl}, that S0 formation preferentially occurs in the group environment. Physical processes which might be relevant, and consequences of this result are outlined in Sections~\ref{sec:disc} and~\ref{sec:concl}. 

Throughout this paper a $\rm \Lambda CDM$ cosmology with 
$\rm \Omega_m =$ 0.3, $\rm \Omega_\Lambda =$ 0.7 and 
$\rm H_0$ = 75$h_{75}\kmsMpc$ is assumed. 
Note that in Section~\ref{sec:disc} the use of data from the literature enforces the consideration of 
different cosmological parameters, which is discussed in detail. 

\section{Data}\label{sec:data}

\subsection{CNOC2 Survey and Groups}\label{sec:cnoc2}

Our sample is based upon the CNOC2 (Canadian Network for Observational Cosmology) redshift survey which is magnitude
limited down to $\Rc$ (Vega) $\sim23.2$ in photometry and totals $\sim$1.5\sqdegr\ within four separate patches on the sky. 
Spectroscopic redshifts, mainly in the range 0.1$\leq z \leq$0.55 exist for a large and unbiased sample of $\Rc \lesssim 21.5$ galaxies \citep{Yee00}. 

Galaxy groups have been identified by applying a friends of friends algorithm to detect significant overdensities in redshift space, with parameters tuned to pick up virialized systems \citep{Carlberg01}: 
In practice this means that group selection is tuned such that at least three galaxies with CNOC2 redshifts are clustered tightly enough that they appear at least 200 times overdense with respect to the critical density.
Whilst incompleteness in the CNOC2 survey leads to incompleteness in the group sample \citep[see][for details]{Carlberg01}, the selection is unbiased and well suited to study the influence of the group environment on galaxy evolution.

\subsection{The ACSR22 Sample}\label{sec:sample}

Supplementary spectroscopy was collected for twenty CNOC2 groups at 0.3$\leq z \leq$0.55 using the LDSS-2 multi-object spectrograph on the Magellan 6.5m telescope to reach an unbiased limit of $\Rc=22$ for spectroscopy and improving the completeness for these systems within 180\arcsec\ (Wilman et al., 2005a\nocite{Wilman05}). We call this the {\bf R22} sample. 
Six serendipitous $z>0.3$ CNOC2 groups in these fields brings the number of R22 groups to 26. 
Deep ($4\times510$~s$=2040$~s) targetted observations of these same 20($+$6) groups with HST-ACS (Hubble Space Telescope - Advanced Camera for Surveys)\footnote{program 9895} with the F775W filter (close to R-band) provide the resolving power and depth to classify morphologies well below the magnitude limit of our spectroscopy. 
This data is processed using the ACS pipeline \citep{HSTACSManualV3}, 
producing a fully astrometrically calibrated and undistorted image from the highly distorted ACS focal surface. 
Other basic calibration procedures which have been applied automatically before we retrieved the data include bias and dark subtraction and flat fielding. %
To create a suitable final image with cosmic rays and hot pixels removed we reapply the \emph{Multidrizzle} task in \emph{pyraf} which recombines the dithered images into an undistorted and well-sampled median image in WCS coordinates. 
The final images look clean and galaxies can be easily identified below the CNOC2 photometric catalogue magnitude limit of $\Rc \sim 23.2$.

Group membership was redefined by Wilman et al., 2005a\nocite{Wilman05} to trace galactic properties further from the group centre.
\footnote{Group membership requires $\Delta(z)\leq 2\times$ the velocity dispersion, and projected distance from the group centre within a limit set such that the aspect ratio of the group in redshift space, b is set to 5 (equation~3 of Wilman et al., 2005a\nocite{Wilman05}).} 
Close examination of the R22 groups reveals significant substructure in many cases. 
This implies an important role for recent accretion onto groups from the surrounding filamentary structure, although group cores may be virialized in most cases. 
Substructure and low number statistics also influence our computed velocity dispersions, which may be overestimated when compared to the stacked weak lensing signal, especially in more massive groups \citep{Balogh07}. 
The ACS field of view is 202\arcsec\ $\times$ 202\arcsec, corresponding to an angular distance of $1.3h_{75}^{-1}Mpc$ at $z=0.55$ and $0.9h_{75}^{-1}Mpc$ at $z=0.3$. 
Centred on a targetted group, this means full coverage with ACS to $0.51h_{75}^{-1}Mpc$ at $z=0.4$, or the virial radius of a relatively massive $\sim550$~km~s$^{-1}$ group at this redshift. 
Taking into account the uncertainty on velocity dispersion, and the un-virialized state of many groups, 
we prefer to keep everything within the R22 sample and with ACS coverage (the {\bf ACSR22} sample), 
rather than applying any additional radial cut. 
The resultant sample of morphologically classified group galaxies is inevitably biased towards members of smaller, more compact groups, sampling only the cores of more massive or loose groups. 
Therefore the sample will suffer little contamination from galaxies unbound to the group 
(which, if they are assigned to groups at all, will tend towards large group-centric radii), 
whilst the influence of group-centric radius on morphological composition will be investigated in Section~\ref{sec:radialtrends}. 
Bias resulting from the small variation of radial limits with redshift within the sample is unimportant, as evolutionary trends at $0.3\leq z \leq0.55$ are not investigated in this paper. 
The ACSR22 sample numbers 179 morphologically classified group galaxies and 111 field galaxies. 
{\bf Field} galaxies are defined to be those galaxies not assigned to any CNOC2 group, 
although some of these may be contained within undetected groups. 
See \citet{McGee08} for a discussion of this topic. 

Deeper CFHT MegaCAM (griz) and CFH12K (BVRI) images are available for most of the CNOC2 area, 
providing more reliable photometry for many galaxies (Balogh et al, in prep). 
To compute total rest-frame r-band AB luminosities for the ACSR22 sample, we use the following procedure. 
We take the $r_{AUTO}$ AB magnitudes from the MegaCAM r-band imaging where available. 
Where this is not available, but CFH12k imaging exists, we correct the CFH12k R-band AB magnitude to the MegaCAM magnitude system using: 
$r_{AUTO}$ (MegaCAM) $=$ $R_{AUTO}$ (12k) $- 0.07 + 0.260990 \times (V_{3}-R_{3})$ (12k)
where the $(V_{3}-R_{3})$ (12k) CFH12K colour is computed using 3\arcsec\ apertures on imaging convolved to a 1\arcsec\ PSF. 
K-corrections to rest-frame r-band ($r_0$) are performed using the \emph{kcorrect} code of \citet{Blanton07kcor}, 
as described by Balogh et al, in prep. 
Out of 290 galaxies in the ACSR22 sample, 49 have no MegaCAM or CFH12K imaging, and so we are forced to rely on the original CNOC2 photometry described by \citet{Yee00}.
Conversion to $M_{r_0}$ (AB) is possible by applying a constant offset of -0.11mag to the rest-frame $\Rc$ (Vega) luminosities derived using the \emph{kcorrect} software, and the CNOC2 $\UBVRI$ photometry. 
For the purposes of this paper we do not need to worry about the residual scatter of $\sim 0.15$mag. 
V-band luminosities (on the CFH12K system) are computed in a similar fashion. 
Total magnitudes are estimated using $V_{AUTO} = R_{AUTO} - (V_3 - R_3)$. 
For galaxies with MegaCAM only photometry, we use: $V_{3}$ (12k) $= G_{3}$ (MegaCAM) $+ .12\times( G_{3}$ (MegaCAM) $-R_{3}$ (MegaCAM) )$^2 - .67\times( G_{3}$ (MegaCAM) $- R_{3}$ (MegaCAM) ) $+.12$, with 0.035mag scatter. 
The V-band luminosity for galaxies with CNOC2 photometry only is simply $M_V$ (12k) $= M_V$ (CNOC2) $- 0.02$, with 0.11mag scatter (good enough for our purposes). 

Weights for the R22 sample have been derived by Wilman et al., 2005a\nocite{Wilman05} to correct for spectroscopic incompleteness. 
We slightly modify these by removing the radial weighting for LDSS-2 galaxies, 
recomputing the spectroscopic selection weight purely as a function of CNOC2 $\Rc$ band magnitude. 
Note that the results presented in this paper are not strongly dependent upon whether or not these or any weights are applied. 

\subsection{Morphological Classifications}\label{sec:class}

The ACS observations reach similar depth to the WFPC2 observations of MORPHS clusters. 
\citet{Smail97} determine that visual classification of $z\sim0.5$ MORPHS galaxies is possible down to a limiting magnitude of $R_{702}\leq23.5$ which corresponds to $\Rc\leq23.9$. 
With the spectroscopic limit of $\Rc = 22.0$, the ACS depth is easily sufficient to achieve robust classification of all ACSR22 galaxies. 

All galaxies in the ACSR22 sample have been visually classified by two of the authors (Augustus Oemler Jr, AO and John Mulchaey, JM), independently from each other. 
Galaxies are visually classified onto the Revised Hubble Scheme in a near identical fashion to the MORPHS sample as described by \citet{Smail97}. 
Comparable classification to the MORPHS sample can be assumed, since AO was one of the MORPHS classifiers.
To briefly describe the classification process: 
each galaxy is displayed as a postage stamp-sized image at 2 scalable stretch levels using a purpose-written \emph{iraf} tool. 
This allows faint features at the galaxy outskirts to be identified as well as bright central features. 
Most classifications should be robust to within $\sim 1$ Hubble Type as in \citet{Smail97}. 
Galaxies are classified in a random order so that the classifier has no prior knowledge of group membership. 
Therefore the comparison between group and field samples is not biased by any prior knowledge of environment by the classifier.\\

\noindent The most difficult classification decisions are:
\begin{itemize}
\item{{\bf E vs S0, for anything other than edge on objects:} 
This relies on seeing a break in the surface brightness gradient (for S0s) and requires a careful look through the entire range of stretches of the image. 
There is a hard outer edge to the bulges of S0s which does not move much as the stretch is changed.}
\item{{\bf S0 vs Sa: } 
The exact dividing line between these types is difficult, and depends on seeing surface brightness irregularities in the disk.}
\end{itemize}

To deal with these cases, the classes E/S0, S0/E and S0/a were used, where the order reflects the preferred class.
Spiral galaxies were classified in half-class increments, and clearly barred galaxies were classified as such. 
Many galaxy types are qualified with \emph{pec} to indicate peculiar features in a galaxy which otherwise fits nicely onto the Revised Hubble Scheme. 
Other than the usual E, S0, spiral and irregular types, other types used were M (merger), X (compact) and ? (galaxies not well described by the Revised Hubble Scheme). 

Classifications by the two classifiers are highly consistent. 
Out of 310 classified galaxies in total, 222 are identical, and of the remaining 88, 69 differ by 0.5 in Hubble type, 17 differ by 1 and just 2 by 1.5. 
The 'final' classification is selected by taking the average hubble type, 
rounding to the nearest full type where there is 0.5 or 1.5 difference. 
In only 2 cases does one classifier select 'Sa' and the other S0/a: on these occasions the Sa classification 
is selected. There is also a single case of E/S0 and S0 classifications: here the S0 classification is preferred. 
Final classifications for a random selection of galaxies are illustrated in Figure~\ref{figure:pstamp} of Appendix~\ref{app:pstamp}. 
It should be noted that these classifications differ slightly from those presented by \citet{Wilman08}, in which classifications from a single author (AO) were used. 

We group galaxies into six broader categories of morphology. 
This avoids oversampling the Hubble sequence, and limits the effects of scatter between Hubble types. 
The broader categories are: {\bf E} (E, E/S0), {\bf S0} (S0/E, S0, S0, S0/a, SB0/a), {\bf early-type spiral, eSp} (Sa-Sbc, including barred spirals), {\bf late-type spiral, lSp} (Sc-Sm, including barred spirals), {\bf irregular, Irr} (Irr, ?) and {\bf merger, M} (M or X). 
However in most of our analysis the spirals are grouped together into a single category.
Peculiar ``\emph{pec}'' qualifications are ignored at this stage. 

To understand the nature of galaxies classified as S0, it is important to recognize that, other than the existence of both bulge and disk components, there is no bulge to total ratio requirement for an S0. 
As understanding the environment of S0s is central to our science, it is important to understand any classification bias or selection effects which might influence what is classified or mis-classified as an S0 (or which should have been classified S0, but was not). 

\begin{figure}
  \epsscale{1.00}
  \plotone{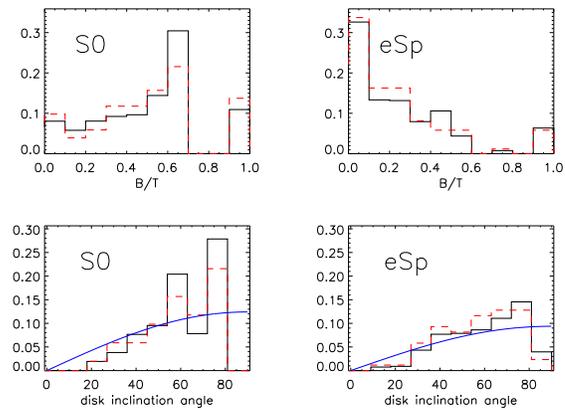}
  \caption{{\bf Top panels:} Solid (dashed) histogram: Distribution of GIM2D-derived bulge to total B/T parameter for visually classified S0 and early-type spiral (eSp) galaxies, weighted (unweighted) for spectroscopic incompleteness. 
{\bf Bottom panels:} As the top panels, but showing the distribution of disk inclination for galaxies with both bulge and disk components. The continuous solid line is simply sin(inclination) and should be expected for a completely random distribution of inclinations.}
  \label{figure:btdiscinctype}
\end{figure}

To address this question, the GIM2D bulge/disk decompositions of \citet{McGee08} provide complimentary parameteric information about the galaxy morphologies. Simultaneous fitting of bulge and disk components provides, amongst other things, bulge to total ratio (B/T), and disk inclination angles (i). 
Figure~\ref{figure:btdiscinctype} shows the distribution of GIM2D derived bulge to disk ratios (B/T), and disk inclination for galaxies which are visually classified as S0s, and eSps. 
Although bulge to disk luminosity does not factor in the decision to call a galaxy S0, it is clear from the top panels that the distribution of B/T is peaked to much larger B/T for S0s than for early type spiral galaxies. 
The sin(i) curve (smooth solid line in the bottom panels) shows what would be expected for a purely random distribution of disk inclination. 
S0 galaxies appear to be slightly biased towards large inclination angles, compared to random. 
This indicates that either the automated fitting biases this parameter, 
or that the visual morphological classification is biased against finding S0 types for low inclination angles.
As it is difficult to identify disks in $\sim$face-on bulge+disk systems with less obvious spiral structure, 
then it is likely that this bias is present in our sample. 
This might lead to an underestimate of the total S0 fraction. 
However we note that some galaxies are fit with pure bulge or disk components, and in these cases the disk inclination angle is not provided \citep[see][]{McGee08}. 
This incompleteness makes any quantification of the S0 fraction underestimate impossible. 
However the estimated fraction should be consistent with other studies which will suffer from the same biases.

\section{Results}\label{sec:results}

\subsection{Composition of Groups}\label{sec:composition}

\begin{figure*}
  \epsscale{0.80}
  \plotone{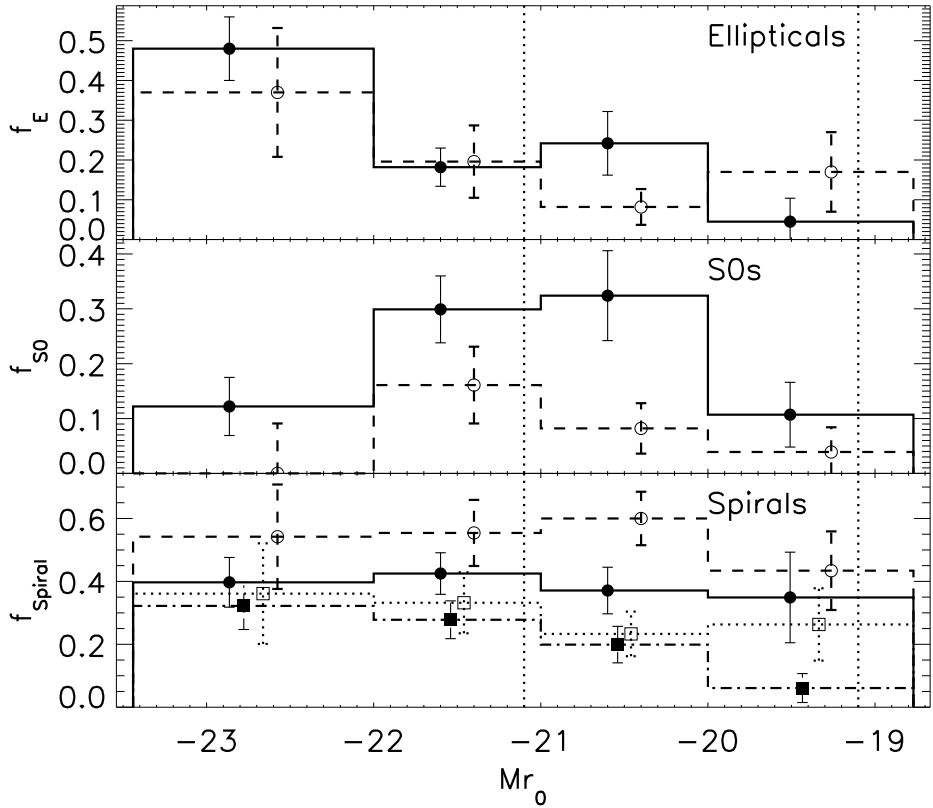}
  \caption{The morphological composition of 0.3$\leq$z$\leq$0.55 CNOC2 group (solid lines and filled points) 
    and field (dashed lines and empty points) galaxies 
    as a function of $r_0$-band luminosity. 
    Redder galaxies will more easily fall out of the sample than bluer galaxies at the same redshift at 
    luminosities below -21.1 at z=0.55 and -19.1 at z=0.3 (vertical dotted lines). 
    Galaxies are weighted to account for spectroscopic incompleteness as a function of $\Rc$-band magnitude.
    The fraction of early type spirals (eSp, Sa-Sbc) is also indicated (squares and dotted/dot-dashed lines).
    Errors on fractions are computed using the Jackknife method and points are offset in luminosity for clarity.} 
  \label{figure:fraccomp}
\end{figure*}

Figure~\ref{figure:fraccomp} shows the fraction of elliptical (E), S0 and spiral (eSp$+$lSp) galaxies in four bins of $r_0$-band luminosity ($<$-22, 1-mag bins down to -20, and a faint bin $>$-20) in both group and field environments. 
The fraction of early-type spirals (eSp) is also indicated (squares).  
It is clearly important to examine these trends not only as a function of environment but also as a function of galaxy luminosity or mass, since this can be as, if not more important to determine a galaxy's properties including morphology \citep[e.g.][]{Blanton03,Kauffmann04,Tanaka04,Baldry06}. 
Indeed the morphological composition of both group and field environments is a strong function of luminosity for our sample. 
Vertical dotted lines indicate luminosity limits at z=0.55 ($\sim$-21.1) and z=0.3 ($\sim$-19.1) given the maximum k-correction of a normal galaxy. 
Fainter than -21.1 lower redshift galaxies will stay in the sample whilst higher redshift galaxies are too faint. 
More importantly, redder galaxies will more easily fall out of the sample than bluer galaxies at the same redshift. 
However this will apply equally to group and field galaxies, and so, qualitatively, the dependence of morphological composition on environment should be robust down to faint luminosities. 
Quantitatively, a higher fraction of red, early type objects might be underestimated due to the more frequent loss of red galaxies. 
This will be a small effect, working only to reduce the significance of our results, and can thus be ignored.  
In the following analysis a luminosity limit of $M_{r_0} = -19$ is applied, excluding only the faintest objects.

Contrasting the group and field compositions, it is surprising to find no highly significant difference in the fraction of ellipticals at fixed luminosity. 

Contrarily, the fraction of S0s is clearly a strong function of environment. 
In groups, S0s contribute 40/178 (weighted fraction 25\%) of all classified galaxies compared to just 10/109 
(weighted fraction 8\%) in the field. 
A Fisher's exact test gives a probability of 0.0005\% (0.0038\% unweighted) that the group and field 
samples are randomly extracted from the same underlying population. 

Galaxies in groups are also biased towards high luminosity with respect to field galaxies. 
This is true in our sample (see Wilman et al., 2005a\nocite{Wilman05}), 
and is consistent with the literature \citep[e.g.][]{Balogh01,Girardi03,DePropris03,Blanton03}. 
To examine the underlying probability distribution more accurately, taking account of the luminosity bias, 
we apply a Monte Carlo method.
This procedure is designed to remove the bias, and test the significance of different morphological 
compositions at fixed luminosity. 

Each group galaxy is selected in turn, and one of the five field galaxies nearest in 
luminosity is randomly selected for a ``mock group'' sample. 
$10^5$ such mock group samples are constructed, and in each one the fraction of S0 type galaxies 
is computed. 
In none of $10^5$ mock groups is the S0 fraction as high as that in the real group sample. 
This is equivalent to the probability that the group and field subsamples are taken from the same underlying 
sample, 
and corresponds to a significant excess of S0s in groups at a confidence level of $>$99.999\% 
($>$4.25$\sigma$ for a one-tailed gaussian distribution, corresponding to the statistical resolution of the resampling process). 
As the group and field samples are roughly equivalent (and relatively small) in number, 
the reverse exercise is also applied to check for consistency. 
i.e. the fraction of mock field samples (constructed using the group sample) with S0 fraction 
as low as the real field sample is computed (in an identical fashion to that described above). 
This results in a confidence level of 99.876\% (2.8$\sigma$) that a deficit of S0s exists in the field sample (compared to the groups). 
Although less significant due to subtle differences, 
this supports the result that there is a excess of S0s in groups, independent of galaxy luminosity. 

To ensure this is not merely a result of including the most massive groups/clusters in the group sample, 
the 2 groups with $\sigma>700$~kms$^{-1}$ (g138 and g38) are excluded. 
This results in 30/145 (weighted fraction 23\%) S0s in the sample, and repeating the bootstrapping procedure 
indicates this still corresponds to a confidence level of 99.995\% (3.9$\sigma$) for the excess. 

The result is also robust against the inclusion of S0/a galaxies in the S0 sample. 
This accounts for 30\% of both field and group S0 populations. 

The same procedure has been applied to show that there is no significant excess of ellipticals in groups 
at fixed luminosity ($\sim$15\% probability that the difference arises randomly) although the absolute 
fraction of ellipticals is much higher, 
with 43/178 (weighted fraction 22\%) compared to 15/109 (weighted fraction 13\%) in the field. 
i.e. This difference is mainly driven by the environmental dependence of the luminosity function 
(a high fraction of the brightest galaxies are ellipticals, see Figure~\ref{figure:fraccomp}). 
The deficit of spiral galaxies in groups exists at a confidence level of 99.953\% ($3.3\sigma$), 
and is seen most clearly in the late-type spiral population (lSp, composed of Sc and later types).

\subsection{Radial Trends}\label{sec:radialtrends}

Whilst it is clear that galaxy properties including morphology must be somehow influenced by the group environment, even at fixed luminosity, there are many ways in which this could be realised. 
To gain further insight into the possible mechanisms involved, it is desirable to see how morphology depends not only on whether a galaxy is a member of a group but also the nature of that group and of the galaxy's local environment. 

One difficulty in studying groups is that one is always limited by the low number statistics for individual systems 
(typically $\sim 10$ members per group). 
Any local density estimate will also correlate strongly with group-centric radius (introduced by the selection of the luminosity-weighted group centre) and number of galaxies in the group (small numbers). 
Therefore to examine trends with local environment within the group, we turn to the group-centric radius. 
Galaxies at small radii will be more subject to processes relating to the global group potential (AGN feedback, merging with the central galaxy), 
whilst galaxies at larger radii are more likely to be influenced by their pre-group environment 
(mergers/interactions in smaller groups, pairs), 
or by processes which are active upon their accretion into the group (stripping of hot gas, harassment).

\subsubsection{Trends and Significance}\label{sec:radialsignif}

\begin{figure}
  \epsscale{1.00}
  \plotone{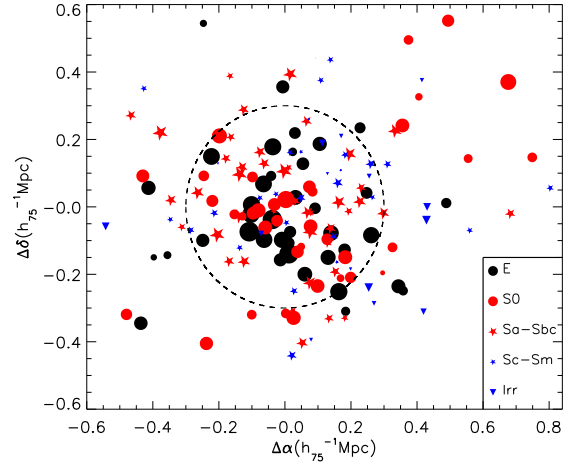}
  \caption{Morphologically classified galaxies in the stacked group sample, to illustrate morphological and luminosity segregation within the groups. Morphological types are indicated by the symbol (see key), and symbols are scaled to a size which is proportional to the galaxy's luminosity. Note that no account is taken for peculiar \emph{pec} type galaxies. The dashed circle at 0.3h$_{75}^{-1}$Mpc is used to divide galaxies into inner group and outer group sub-samples. These are compared in Table~\ref{tab:radialfracs} and in the main text to assess the significance of radial trends.}
  \label{figure:stackedgrp}
\end{figure}

\begin{table*}
\begin{center}
\caption{Morphological composition of CNOC2 group and field samples ($0.3\leq z\leq 0.55$). Composition (fractions, weighted to account for spectroscopic incompleteness) of total group, inner ($<0.3h_{75}^{-1}Mpc$) and outer ($\geq0.3h_{75}^{-1}Mpc$) groups, and the field are computed for 3 ranges in galaxy luminosity. (Unweighted) Number of galaxies of each type are indicated inside the brackets. Errors are computed using a Jackknife method, except for N$\leq1$, for which a generic resolution error of $\frac{1.}{N_{all}}$ is indicated. Note that these symmetric errors are a rough guideline only, although fractions of $<$0 and $>$1 are clearly unphysical. More robust estimates of the significance of environmental trends are described in the main text.}
\label{tab:radialfracs}
\vspace{0.1cm}
\begin{tabular}{cccccc}
\hline\hline
\noalign{\smallskip}
Type & Luminoisity range & groups & $<0.3h_{75}^{-1}Mpc$ & $\geq0.3h_{75}^{-1}Mpc$ & field \\
E & -23.5$\leq M_R \leq$-19.0 & 22$\pm$4\%(43) & 21$\pm$4\%(28) & 24$\pm$7\%(15) & 13$\pm$4\%(15)\\
  & -23.5$\leq M_R \leq$-21.0 & 28$\pm$4\%(33) & 33$\pm$6\%(26) & 18$\pm$6\%(7) & 24$\pm$8\%(9)\\
  & -21.0$\leq M_R \leq$-19.0 & 18$\pm$6\%(10) & 5$\pm$4\%(2) & 27$\pm$9\%(8) & 10$\pm$4\%(6)\\
S0 & -23.5$\leq M_R \leq$-19.0 & 25$\pm$4\%(40) & 20$\pm$5\%(22) & 30$\pm$7\%(18) & 8$\pm$3\%(10)\\
   & -23.5$\leq M_R \leq$-21.0 & 24$\pm$4\%(25) & 24$\pm$6\%(17) & 24$\pm$8\%(8) & 12$\pm$5\%(5)\\
   & -21.0$\leq M_R \leq$-19.0 & 25$\pm$7\%(15) & 14$\pm$7\%(5) & 33$\pm$10\%(10) & 7$\pm$3\%(5)\\
eSp & -23.5$\leq M_R \leq$-19.0 & 21$\pm$3\%(47) & 26$\pm$5\%(29) & 16$\pm$4\%(18) & 27$\pm$5\%(31)\\
    & -23.5$\leq M_R \leq$-21.0 & 29$\pm$5\%(32) & 31$\pm$6\%(22) & 26$\pm$8\%(10) & 34$\pm$8\%(14)\\
    & -21.0$\leq M_R \leq$-19.0 & 15$\pm$4\%(15) & 20$\pm$8\%(7) & 12$\pm$5\%(8) & 26$\pm$6\%(17)\\
lSp & -23.5$\leq M_R \leq$-19.0 & 17$\pm$4\%(30) & 20$\pm$5\%(18) & 15$\pm$5\%(12) & 27$\pm$5\%(31)\\
    & -23.5$\leq M_R \leq$-21.0 & 12$\pm$3\%(13) & 11$\pm$4\%(8) & 14$\pm$6\%(5) & 21$\pm$8\%(8)\\
    & -21.0$\leq M_R \leq$-19.0 & 21$\pm$6\%(17) & 30$\pm$11\%(10) & 15$\pm$7\%(7) & 28$\pm$6\%(23)\\
Irr & -23.5$\leq M_R \leq$-19.0 & 14$\pm$4\%(17) & 14$\pm$6\%(7) & 14$\pm$5\%(10) & 24$\pm$5\%(21)\\
    & -23.5$\leq M_R \leq$-21.0 & 6$\pm$3\%(4) & 0$\pm$1\%(0) & 19$\pm$9\%(4) & 9$\pm$4\%(4)\\
    & -21.0$\leq M_R \leq$-19.0 & 20$\pm$6\%(13) & 32$\pm$12\%(7) & 12$\pm$5\%(6) & 28$\pm$7\%(17)\\
M & -23.5$\leq M_R \leq$-19.0 & 0$\pm$1\%(1) & 0$\pm$1\%(0) & 1$\pm$1\%(1) & 2$\pm$1\%(1)\\
  & -23.5$\leq M_R \leq$-21.0 & 0$\pm$1\%(0) & 0$\pm$1\%(0) & 0$\pm$3\%(0) & 0$\pm$3\%(0)\\
  & -21.0$\leq M_R \leq$-19.0 & 1$\pm$1\%(1) & 0$\pm$3\%(0) & 1$\pm$3\%(1) & 2$\pm$1\%(1)\\
\noalign{\hrule}
All & -23.5$\leq M_R \leq$-19.0 & (178) & (104) & (74) & (109)\\
    & -23.5$\leq M_R \leq$-21.0 & (107) & (73) & (34) & (40)\\
    & -21.0$\leq M_R \leq$-19.0 & (71) & (31) & (40) & (69)\\
\noalign{\hrule}
\end{tabular}
\end{center}
\end{table*}

All group galaxies in the ACSR22 sample are compiled as a function of their physical group-centric offsets $\Delta\alpha$ and $\Delta\delta$ into a \emph{stacked group}, illustrated in Figure~\ref{figure:stackedgrp}. 
Galaxies are keyed by their morphologies, with symbol size proportional to galaxy luminosity to illustrate the morphological and luminosity segregation within our groups. 
To highlight the influence of group-centric radius on morphological composition, 
the fraction of different types is computed independently for group members at distances $<0.3h_{75}^{-1}Mpc$ 
and $\geq0.3h_{75}^{-1}Mpc$ from the group centre. 
At $z=4$, this corresponds to the virial radius of a group with velocity dispersion $\sim320\kms$. 
The dashed circle in Figure~\ref{figure:stackedgrp} corresponds to this division. 
These, along with total group and field fractions, are summarized in Table~\ref{tab:radialfracs}.
Fractions are weighted to account for spectroscopic incompleteness, but the total number of galaxies for each 
type is also given in brackets, and the total number of galaxies is provided for Type$=$All. 
All this information is provided for 3 luminosity bins: 
combined (-23.5$\leq M_R \leq$-19.0); bright (-23.5$\leq M_R \leq$-21.0); and faint (-21.0$\leq M_R \leq$-19.0). 
This allows trends to be investigated separately for bright and faint galaxies, which are likely to 
experience different physical processes \citep[e.g.][]{Tanaka04}. 

Some differences between the composition of inner and outer group are immediately apparent in Table~\ref{tab:radialfracs}. 
The significance of these differences is assessed as in Section~\ref{sec:composition}: 
i.e. by computing the probability that the fraction of a given type in the inner group can be reproduced 
by sampling from the outer group population, matching the galaxies on luminosity to remove that dependence 
(since there is a strong correlation of luminosity with group-centric radius, such that the 
brighter galaxies tend to reside near the group centre). 
The enhancement of bright ellipticals in the inner group is highly significant 
($<0.001\%$ probability that it could be randomly extracted from the outer group population), 
whilst the significance of the faint elliptical fraction increasing with radius, as seen in Table~\ref{tab:radialfracs}, disappears with luminosity-matching: i.e. this may be purely a consequence of the luminosity-radius trend for fainter galaxies, folded with the luminosity distribution of faint ellipticals. 
S0s are altogether more common in the outer group, 
but only at the 
97\% confidence level, 
less significant for faint or bright sub-samples. 
Late-type spirals are significantly more common in the inner group 
(99.85\% confidence level). 
This appears to be driven mainly by the fainter galaxies, and drives the overall trend in spiral galaxies. 
Early-type spirals on the other hand live in environments consistent with those of other galaxies of equivalent luminosity. 
This interesting result, in apparent contradiction with expectations, originates from the unavoidable correlation between group size and the average distance of group members from the centre, and from the concentration of faint late-type spirals in the smallest groups in the sample. 

Otherwise, the morphological composition shows no strong correlation with velocity dispersion within our sample, 
as noted for the fraction of passive (EW[OII]$\lambda3727 <$5\AA) galaxies (Wilman et al., 2005a\nocite{Wilman05}). 
This is influenced by the small number statistics and uncertainties in velocity dispersion which hinder studies of individual groups. 
However the significant radial trends discussed above are not significantly effected if the 2 groups with $\sigma>700$~kms$^{-1}$ (g138 and g38) are excluded. 

\subsubsection{Uncertainty of Group Centres}\label{sec:groupcenuncert}

Radial relations are of course subject to the uncertainty on the computed group centres, 
which are simple luminosity-weighted centres computed iteratively as described by Wilman et al., 2005a\nocite{Wilman05}, 
picking out the dense core as shown in Figures~4 and~5 of that paper. 
The existence of strong radial segregation of the galaxy population provides independent evidence that 
these centres are physically meaningful, with an accuracy of $<<$0.3h$_{75}^{-1}$Mpc. 
That the morphological segregation is independent of luminosity, and is therefore independent of any 
parameter involved in the definition of the group centres, means that this is a robust result. 

\begin{figure}
  \epsscale{1.00}
  \plotone{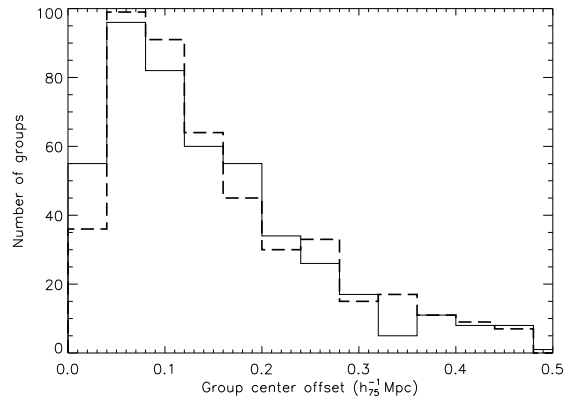}
  \caption{Distribution of offsets for the observationally-defined group centre from the true centre 
    of the mock catalogue of friends-of-friends groups \citep{McGee08}. The observationally defined 
    centre is the luminosity-weighted centre of randomly sampled galaxies to simulate incompleteness 
    in the data. The true centre is defined to be the position of the central galaxy of the most massive 
    halo (thick, dashed line) or the luminosity-weighted centre of all galaxies (incompleteness$=$0, thin line).}
 \label{figure:mockgroupoffset}
\end{figure}

To examine the effect of group centre uncertainty, we examine the distribution of expected offsets using the 
mock catalogue of friends-of-friends groups generated by \citet{McGee08}. 
Figure~\ref{figure:mockgroupoffset} shows the distribution of group centre offsets computed as observationally 
(luminosity-weighted centre of randomly sampled galaxies to simulate incompleteness in the data) from the 
best-estimate of the true centre. 
The true centre is chosen to be position of the central galaxy of the most massive halo (thick, dashed line, 
where in the semi-analytic model this is by definition at the halo centre), and the luminosity-weighted 
centre of all galaxies (incompleteness$=$0, thin line). 
The mean (median) offsets from the central galaxy are 0.15h$_{75}^{-1}$Mpc (0.13h$_{75}^{-1}$Mpc), and from the luminosity-weighted 
centre of all galaxies are 0.14h$_{75}^{-1}$Mpc (0.12h$_{75}^{-1}$Mpc). 

An uncertainty of 0.15h$_{75}^{-1}$Mpc in the group centre can be folded into the Monte Carlo procedure 
by allowing the centre to vary by a random amount sampled from a gaussian with width $\sigma=$ 0.15h$_{75}^{-1}$Mpc 
on each separate iteration, recomputing the fractions each time. This is folded into the usual resampling 
procedure as described, including luminosity-matching. 
Even with the application of this additional uncertainty, the bright elliptical population is still 
quite significantly (at the 97.5\% confidence level) higher in the inner group ($<$0.3h$_{75}^{-1}$Mpc). 
Of the other trends, the most significant remaining trend is the excess of faint late-type spirals in the 
inner regions (at the 93.4\% confidence level). 
These results are robust considering that the average centre shift is half the magnitude of the radial division, 
which inevitably leads to reduced significance in radial trends. 
Indeed, the latter estimates of significance are highly conservative, and are measured with the premise that 
not only are galaxies within 0.3h$_{75}^{-1}$Mpc of the position chosen to be the centre different from those at 
larger distances (at fixed luminosity), but that those centres are the true centres of the groups. 
If the centres were already wrong, this procedure will only make them more wrong 
(by typically $\sqrt(2)\times$0.15h$_{75}^{-1}$Mpc$ = \sim$0.21h$_{75}^{-1}$Mpc). 
Therefore it is remarkable that any significant segregation still exists in these groups. 
This is a direct indication that morphological segregation (independent of luminosity) and luminosity segregation 
itself (maximized due to the choice of luminosity-weighted centres) 
trace the same underlying physical structures and sub-structures (the dark matter distribution). 

\section{Discussion}\label{sec:disc}

\subsection{S0 formation: Evolution and Environment}

\begin{figure}
  \epsscale{1.00}
  \plotone{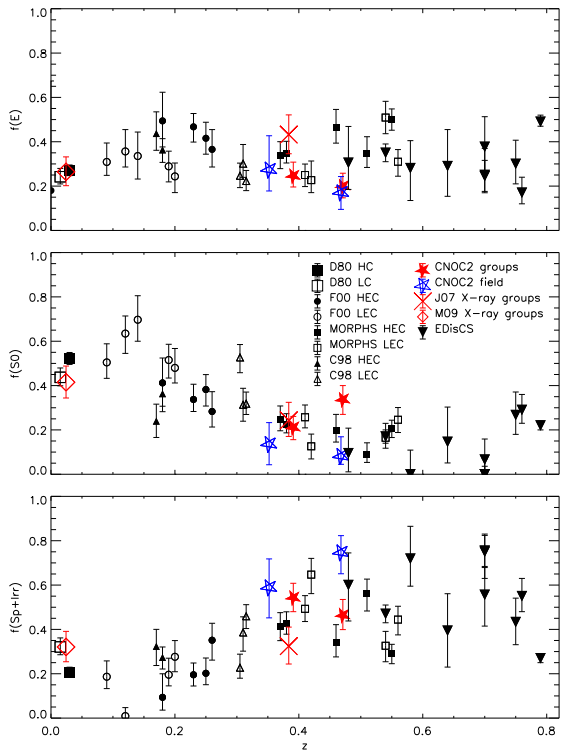}
  \caption{The fraction of elliptical, S0 and spiral$+$irregular types as a function of redshift for galaxy clusters, reproduced from Figure~9 of F00. Overplotted at the median galaxy redshifts are the values from the CNOC2 group and field sample down to the luminosity limit of F00. Also plotted are the values for the z$=$0.23-0.6  X-ray group sample of J07, at the median group redshift, the local X-ray group sample of M09, and the individual EDisCS clusters (D07). For more details of selection, cosmology and error computation, see the main text.}
 \label{figure:ftypez}
\end{figure} 

We return to the main question asked by this paper: 
Can the observed increase in cluster S0 populations since $z \sim 0.5$ be driven by infall from a group environment in which S0 formation is prevalent? 

We approach this question by simultaneously examining the morphological composition of clusters, groups and the field at different redshifts, starting with the cluster data presented in Figure~9 of F00 (Fasano, private communication). 
In that figure, F00 present the fractions of elliptical (E), S0 and spiral$+$irregular galaxies as a function of redshift for galaxy clusters, selected from their own data ($z\sim 0.2$) and from D80 (low z), MORPHS (D97,$z\sim 0.4-0.6$) and C98 ($z\sim 0.2-0.35$). 
Low redshift D80 clusters are divided into high and low concentration (HC/LC) samples. 
At intermediate and high redshift, F00 instead divide the sample into two samples on the basis of the central concentration of elliptical galaxies (high: HEC and low: LEC), which they show correlates very well with the global ratio of S0s to ellipticals. 
We extend this dataset in Figure~\ref{figure:ftypez} to include galaxies in groups and the field at intermediate redshift. 
F00 data is reproduced using their symbols (corrected for spatial incompleteness, classification bias and field contamination as described by F00). 
We replace their simple Poissonian errors with a better statistical approximation of assymetric errors which combines the Poissonian probability of total number observed with the binomial probability of observing a particular number of a given morphological type. These are derived from equations (21) and (26) of \citet{Gehrels86}. 
As claimed by F00 and D97, there is a strong evolution in the fraction of S0 galaxies (increasing to low z) and spiral$+$irregular galaxies (decreasing to low z), whilst the elliptical fraction remains roughly constant. 
Indeed, there is no evidence of any increase in the elliptical fraction of cluster galaxies since $z\sim1$ \citep[D97; F00;][]{Desai07,Smith05,Postman05}, which is unsurprising if ellipticals are the result of major mergers which are not thought to be common in massive clusters due to the high pairwise velocity of galaxies. 
Interestingly, the evolution in cluster S0 fraction is only seen at $z\lesssim0.5$, with higher redshift studies suggesting a flattening to at least $z\sim1$ \citep[see ][]{Smith05,Postman05}. 

The E, S0 and spiral$+$irregular fractions are computed and overplotted in two redshift bins for CNOC2 group and field samples ($0.3\leq z\leq0.55$, with the sample divided at the median redshift of $z=0.43842$: stars at the median galaxy z for each bin). 
We reiterate that the field sample is likely to contain some galaxies in undetected groups \citep[see][]{McGee08}. 
Also shown are the fractions for X-ray selected groups at $0.23 \leq z \leq 0.6$ from J07 (cross at the median group z), 
a low redshift sample of X-ray selected groups in the Sloan Digital Sky Survey (SDSS) \citep[][M09]{Mulchaey09}, 
and individual higher redshift clusters/massive groups from the EDisCS survey \citep[][D07,$0.58\leq z \leq 0.79$]{Desai07}. 
Whilst EDisCS galaxies are morphologically classified by different persons, they claim 
that, by training using the visual morphological catalogues of the MORPHS clusters and the same classification procedure, they conform to the MORPHS system. 
In all cases, errors are computed using equations (21) and (26) of \citet{Gehrels86} (as already applied to EDisCS clusters by D07). 

\subsection{Selection and Systematics}

It is necessary to consider the different criteria used for galaxy selection in Figure~\ref{figure:ftypez}, 
although it is clearly impossible to mimic selection for groups and clusters of different mass, size, substructure properties and redshift. 
There will also be intrinsic variation in the galaxy mass functions of different systems.
The limiting luminosity applied by F00 is $M_V=-20.0$.  
However this is computed for a $\Omega_m=1$, flat Universe with $H0=50kms^{-1}Mpc^{-1}$. 
The same limit is applied to EDisCS clusters in the same cosmology (D07, Table~3). 
Therefore for CNOC2 groups and field, and for J07 groups, an equivalent luminosity for $z=0.4$ in our assumed $\Lambda$CDM Universe with $H0=75kms^{-1}Mpc^{-1}$ of $M_V=-20.53$ is applied. 

It is not enough to compute luminosity limits in a consistent cosmology. 
D97 claim that the D80 sample actually only reaches $M_V=-20.4$ at $z=0.04$. 
This corresponds to $M_V=-21.23$ in our assumed cosmology (which is more realistic, given modern constraints). 
Therefore there is a difference in luminosity limit of $0.7$mags compared to $z=0.4$ (and more to higher z). 
The CNOC2 groups and field data both show a tendency to lower S0 fraction and higher E fraction for brighter luminosity limits (see Figure~\ref{figure:fraccomp}). 
If the cluster data follows the same trend (expected, since the brightest galaxies are usually ellipticals), 
then the brighter limit used at low z infers that the S0 fraction is underestimated. 
If this is so, then the expected evolution should be even stronger than observed. 
Note however that early-type galaxies are expected to fade by $\sim0.63$mags in V-band between $z=0.4$ and $z=0$ 
\citep[][4Gyr old population at $z=0.4$, solar metallicity, Chabrier IMF]{BC03}, 
effectively cancelling out the effects due to incorrect cosmology and luminosity limit for galaxies which were already passive at $z=0.4$ (by definition not all, given the observed evolution). 
For a low redshift comparison, M09 group sample fractions are also computed down to the D80 limit. 

Spatial completeness is an even trickier selection issue than luminosity. 
F00 compute fractions for each cluster within the available area (0.4-1.8h$_{50}^{-1}$Mpc$^2$, Table~2 of F00), 
although a statistical correction for incompleteness due to the irregular shape of the surveyed
area has been applied (see Sections 4 and 4.1 of F00). 
It is clear from Figure~6 of F00 that the ratio of S0s to ellipticals increases more quickly with radius for HEC clusters 
than for LEC clusters, and thus the offset between these two types might relate to the relative cluster area 
sampled. 
However this effect might be less important than the intrinsic cluster to cluster variation, 
and variable cluster sizes and levels of substructure make it impossible to make a completely fair comparison (see discussion below). 
Fractions are computed for EDisCS clusters within a constant aperture of 600h$_{50}^{-1}$kpc, 
and the large scatter between clusters cannot be attributed to the cluster mass or size: 
the two clusters with N(S0)/N(E)$\geq 0.89$ (admittedly with large error bars) are of intermediate velocity 
dispersion for that sample ($500-650$kms$^{-1}$). 
Spatial selection for groups is discussed in section~\ref{sec:sample} for CNOC2 and by J07 for their sample. 
M09 select all group galaxies within 1h$_{75}^{-1}$Mpc of the group centre. 
This is not sensitive to radial selection: 
inspection shows that an almost identical S0 fraction exists within 500h$_{75}^{-1}$kpc. 
Spatial selection in the field is meaningless in the context of spatial segregation. 

It is clearly impossible to create a perfectly equivalent set of clusters or cluster galaxies, 
given the intrinsic variance in properties and formation histories. 
Adding the redshift or environment axes simply adds to the complexity of this comparison, 
introducing amongst other things a variation of the galaxy mass function on redshift and environment. 
Statistical errors are relatively large, since the luminosity limit applied here samples only the bright end of the luminosity function. 
Other potential errors include classification bias, as discussed by \citet[][]{Andreon98} and \citet[][]{Fabricant00}. 

Despite these many problems, the strong, average evolution of the S0 fraction for bright galaxies in clusters is compelling, with a low f(S0)$< 0.3$ persisting in clusters taken from many different datasets at $z\gtrsim0.35$. 

\subsection{S0 Formation Mechanisms and Implications}

As we have seen, the S0 fraction in optically selected CNOC2 groups is at least as high as that seen in clusters at the same redshift. 
This provides critical evidence that S0 formation is not only possible, but common outside of rich clusters. 
Formation of S0s in groups must continue down to z$\sim$0: whilst we have no optically selected sample of low redshift groups, the z$\sim$0 X-ray selected sample of M09 contains a fraction of S0s similar to that in the D80 cluster sample. 
Thus, the evolution in group S0 fraction likely parallels the evolution in star forming fraction \citep[e.g. Wilman et al., 2005b\nocite{Wilman05b};][]{Poggianti06}.
This contrasts with the lack of evolution in the elliptical fraction which remains low at all redshifts, even in clusters. 
Bright ($M_V\lesssim-20.53$) ellipticals exist with similar frequency in all environments, although they are more likely to be found towards the centre of halos. 

Our result - that S0s are more common in groups than in less dense field environments - implies that environment is a relevant factor in their formation, and that internal secular processes \citep[in any case expected to lead to the formation of star forming pseudobulges, e.g.][]{Kormendy04} may be less important. 
However, since S0s are at least as common in less massive optically selected groups as they are in $z\sim0.4$ X-ray bright groups and clusters, mechanisms which take place in lower density regimes must be active. Even in groups, the mildly significant increase of S0 fraction beyond $0.3h_{75}^{-1}Mpc$ (with 97\% confidence) hints at formation in even lower density regimes such as the group outskirts, or pre-processing in galaxy pairs and lower mass groups.
The growth at late times of the S0 population, in all measured environments, suggests that their formation is a slow process. 
This is consistent with spectro-photometric properties of S0 galaxies in infalling groups and cluster outskirts at $z\sim 0.5$ \citep{Moran07}. 
Finally the early-type spiral population is approximately consistent with that seen in the field, at fixed luminosity, whilst there is a clear deficit of late (Sc+) type spirals. 
This is consistent with the result of \citet{Poggianti08} that the relationship between spiral fraction and density in EDisCS clusters is driven by Sc+ type spirals. 

Converting late-type spiral galaxies into S0s is not a straightforward business, as it involves significant bulge growth even at fixed luminosity in order to match the B/T distribution of S0s (Figure~\ref{figure:btdiscinctype}). 
However the total luminosity of many z$\sim$0.4 spirals is sufficient for them to be progenitors of a typical group S0 (see Figure~\ref{figure:fraccomp}). 
Bulge growth may or may not be related to the actual suppression of star formation and spiral arms in S0s, but the two processes are certainly correlated. 
If bulge growth is achieved via formation of new stars, this puts a lower limit on the level of star formation required. 
Let us assume an extreme case in which a z$=$0.4 pure disk galaxy (B/T$=$0) will become a massive S0 galaxy with stellar mass $2\times10^{11}\Msol$ and $B/T=0.7$ by z$=$0. 
This corresponds to a total bulge mass growth of $1.4\times10^{11}\Msol$ within $\sim4\Gyr$ which requires a bulge star formation rate of SFR$\sim35\Msolyr$. 
A significant population of infrared-bright spiral galaxies exists at 0.4$\lesssim$z$\lesssim$1, for which star formation rates of $\sim 10-100\Msolyr$ have been derived \citep{Bell05,Geach06,Bai07,Geach08,Koyama08}. 
Most of this star formation is obscured at optical wavelengths \citep{Bai07,Geach08}, and these galaxies tend to exist on the outskirts of clusters \citep{Bai07,Marcillac07,Geach08,Koyama08}, and in intermediate density environments, corresponding to groups and sub-groups \citep{Marcillac08}. 
At lower redshifts, optically red 24\micron-emitting galaxies still prefer intermediate density environments \citep[e.g.][]{Gallazzi08}. 
We have already identified a population of galaxies with low [OII] emission but which appear to be star forming at infrared wavelengths \citep{Wilman08}. Preliminary analysis of 24\micron\ data shows that these galaxies frequently form stars at rates $\gtrsim10\Msolyr$ (Tyler et al, in prep). Therefore the data do support a mode of bulge growth involving star formation which is most easily visible at infrared wavelengths, and which prefers group and sub-group environments. 

In these relatively low density environments, interactions with the Intra-Group Medium (IGM), including a gentler strangulation/starvation of the galaxies' hot gas halo, are not likely to be as important as they would be in the group cores, X-ray groups and clusters. 
To promote slow bulge growth in low density environments, we suggest that minor mergers \citep[supported by the existence of kinematically decoupled young cores, e.g.][]{McDermid06}, galaxy harassment \citep[e.g.][]{Moore99}, and tidal interactions between galaxies are more likely to be relevant \citep[see][for more discussion]{Moran07}. 
Each of these mechanisms is capable of funnelling material to the central bulge over a long timescale, and so a massive starburst is not required. 
Harassment and tidal interactions in particular might also promote an infrared-bright mode of star formation \citep{Marcillac08}, although neither mechanism has yet been investigated in detail at the group scale. 
However in each case it is unclear how star formation is eventually suppressed: possibly there is a role for feedback from AGN, given that black hole growth should accompany bulge growth. 
Indeed, bulge growth in low density environments should be expected, since it is typical of the environment in which high-accretion, optically selected AGN are known to reside \citep[e.g.][]{Kauffmann04,Shen07}.

Finally, a dominant route to the formation of S0 galaxies in the group or sub-group environment has important implications for galaxy evolution studies: 
Most galaxies in the Universe are experiencing a group like environment \citep[e.g.][]{Eke04} in which S0 formation processes might be important. 
This means that the quenching of star formation associated with the formation of an S0 might contribute significantly to the global decline in star formation \citep[e.g.][]{Lilly96,Madau98}. 
Pre-processing in groups is also likely to dominate the production of cluster S0s, helping to explain the lack of any strong relationship between S0 fraction and cluster-centric radius or local density (e.g. D80,D97,F00). 
This might still be supplemented by the formation of some S0s in more isolated environments by secular processes (i.e. there are some field S0s in our sample, although note the definition of field in Section~\ref{sec:sample}),
 or by stripping processes in clusters \citep{Moran07}. 
The distribution of B/T for visually classified S0s as a function of environment might help disentangle these different routes. 
The end result for clusters is that the increasing S0 fraction and evolving morphology-density relation (including a declining spiral fraction and stronger relations at low z for low concentration clusters (D97) and groups \citep{Postman84}) is imprinted by the accumulated accretion history of pre-processed galaxies onto that cluster.

\section{Conclusions}\label{sec:concl}

With high resolution ACS imaging of our well studied sample of optically selected galaxy groups from the CNOC2 galaxy survey at $0.3\leq z\leq0.55$, we have been able to visually classify the morphologies of 179 spectrally confirmed group galaxies and 111 field galaxies in this redshift range. 
This provides good enough statistics to examine the morphological composition of galaxies in this important environment at $z\sim0.4$. 

For the first time, we directly address the question of whether the strong increase of S0 fraction in galaxy clusters since $z\sim0.4$ is a result of pre-processing in groups and lower density environments. 
Within our sample, there is a much higher fraction of S0s in groups than in the lower density field at fixed luminosity, that cannot be reproduced by resampling the field population with $>$99.999\% confidence. 
There is also a hint that S0s in groups are less common within 0.3h$_{75}^{-1}$Mpc of the group centres than at larger distances (97\% confidence). 

In contrast to the situation for S0s, the fraction of elliptical galaxies at fixed luminosity is not significantly enhanced in groups compared to the field. 
However there is a strong ($>$99.999\% confidence) enhancement of bright ($M_R \leq$-21.0) ellipticals within 0.3h$_{75}^{-1}$Mpc of the group centres. 
This result is robust against uncertainty in the estimated group centres. 
The presence of bright central elliptical galaxies in many groups suggests that the luminosity weighted centres are indeed good, on average, to $\sim$ 0.15h$_{75}^{-1}$Mpc (as found in mock catalogues). 

Spiral galaxies make up the majority of remaining galaxies, with a strong suppression in groups, compared to the field (99.953\%), mostly late (Sc+) type spirals. 
It is also interesting that (particularly faint, -21.0$<M_R\leq$-19.0) late-type spirals are more common in the inner group regions ($<$0.3h$_{75}^{-1}$Mpc) than at larger group-centric radii (with a confidence level of 99.85\%). 

We have compared the S0 fraction computed for the CNOC2 groups and field samples with a compilation of clusters over the redshift range $0\lesssim z \lesssim 0.8$ (F00,D80,D97,C98,D07) and X-ray selected groups at $z\sim0.4$ (J07) and $z\sim0$ (M09). 
S0s are present with at least as high (and possibly enhanced) fractions in optical groups with respect to all other environments, whilst the fraction in X-ray groups is consistent with that in clusters from $z\sim0$ to $z\sim0.4$. 
Together with the lack of any strong radial trend in groups or clusters, this paints a picture in which low mass groups are likely to be the dominant environment for the formation of S0s, leading to an evolution in the global and cluster S0 fraction as more S0s are formed and accreted onto clusters by the present day. 
The combined evidence supports a slow acting mechanism for S0 formation in which starbursts are not usually evident and many S0s are formed at late times ($z\gtrsim0.5$). 
We emphasize the importance of bulge growth to match the B/T distribution of S0s, and show that measured star formation rates of infrared bright spiral galaxies in similar environments would be sufficient over a few gigayears.
The favoured candidates are therefore minor mergers, galaxy harassment and tidal interactions. 
Mechanisms requiring interaction with a bright X-ray emitting IGM are ruled out as the likely route to formation for the majority of S0s in the Universe.\\

\acknowledgements

We would like to thank the CNOC2 team for allowing us access to their
unpublished data. This work is based on
observations made with the NASA/ESA Hubble Space Telescope, at the
Space Telescope Science Institute, which is operated by the Association
of Universities for Research in Astronomy, Inc., under NASA contract
NAS 5-26555. These observations are associated with program 9895. 
We also extend special thanks to Lori Lubin for the iraf tool used to 
classify galaxies, and to Gionvanni Fasano, Tesla Jeltema and Bianca Poggianti 
for their data and help in constructing Figure~\ref{figure:ftypez}. 
Thanks also to the anonymous referee, and to 
Peter Erwin, Gabriella de Lucia, Jim Geach and Sean Moran for useful and 
interesting discussions which led to the improvement of this paper.
J.S.M. acknowledges partial support from NASA grant NNG 04-GC846 and HST grant G0-09895.01-A.

\appendix

\section{Postage Stamp Images of Galaxies}\label{app:pstamp}

\begin{figure*}
  \epsscale{0.75}
  \plotone{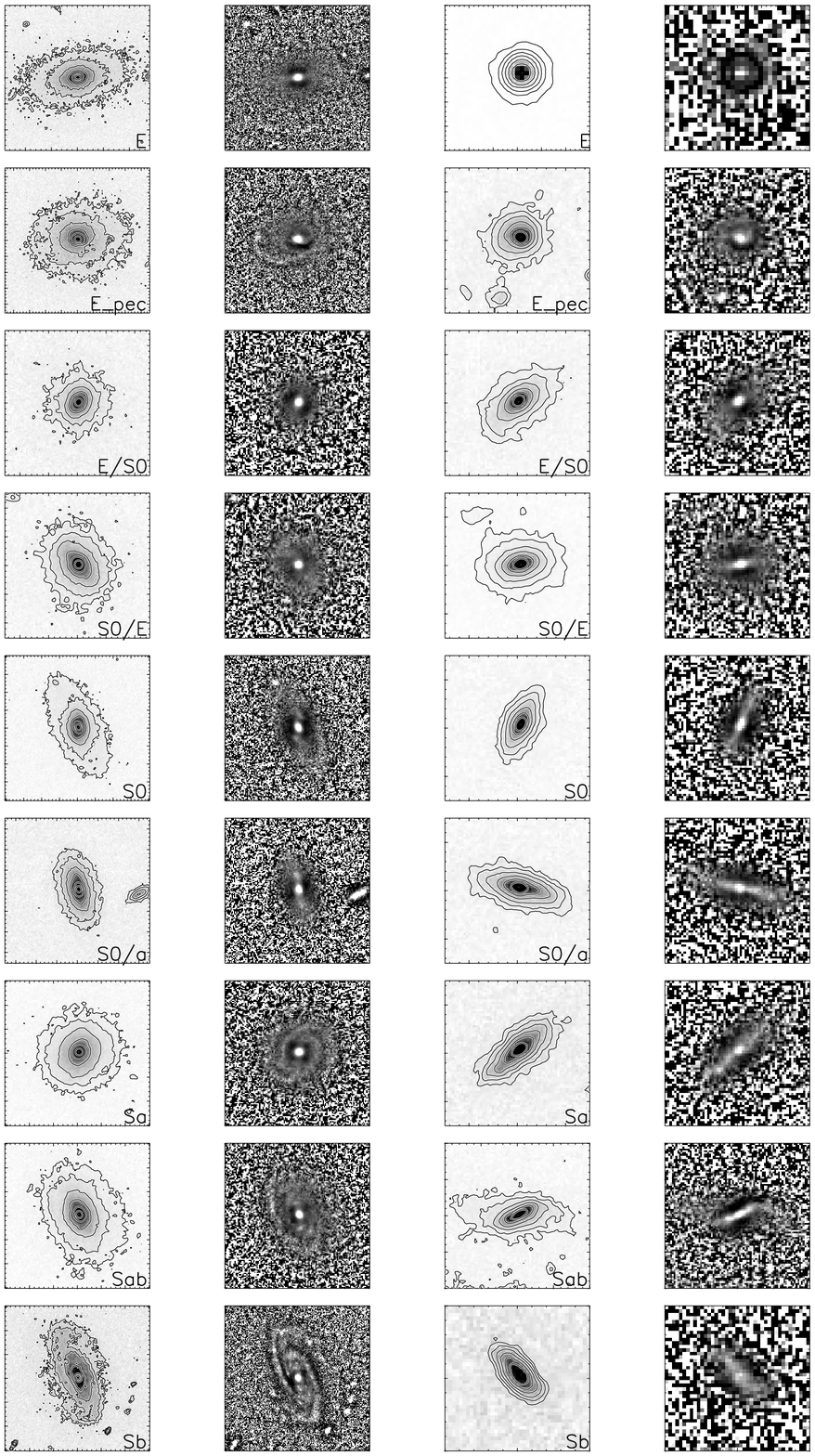}
  \caption{A representative sample of galaxy images.
    One bright (2 left columns) and one faint (2 right columns) galaxy is randomly selected for each of: 
    E, Epec, E/S0, S0/E, S0, S0/a, Sa, Sab and Sb visual classifications. 
    Each galaxy is displayed using linear scaling with log-contours overplotted (left), 
    and using the unsharp mask method (right) to bring out the assymetrical and disk features.
  }
 \label{figure:pstamp}
\end{figure*}

Figure~\ref{figure:pstamp} illustrates the image quality of the ACS data, which makes it suitable for visual classification of morphologies across the full range of Hubble type and luminosity. 
A random selection of galaxies are displayed both using a linear scaling with log-contours, 
which illustrates the low surface brightness features, 
and using the ``unsharp mask method'', which brings out irregularities in the galaxy profile. 
This is implemented by dividing the image at full resolution by a smoothed version of itself.

\end{document}